\documentclass[onecollarge,natbib]{svjour2}
\bibpunct{[}{]}{;}{n}{}{,} 
\smartqed  
\usepackage{graphics}
\usepackage{graphicx}
\usepackage{epsfig}

\usepackage{amssymb}
\usepackage{amsmath}

\journalname{Few-Body Systems}
\begin{document}

\title{Effective field theory of interactions on the lattice
}


\author{Manuel Valiente         \and
        Nikolaj Thomas Zinner 
}


\institute{Manuel Valiente \at
              SUPA, Institute for Photonics and Quantum Sciences\\ 
              Heriot-Watt University\\ 
              Edinburgh EH14 4AS\\
              United Kingdom \\
              \email{M.Valiente\_Cifuentes@hw.ac.uk}           
           \and
           Nikolaj Thomas Zinner \at
              Department of Physics \& Astronomy\\ 
              Aarhus University\\
              DK-8000 Aarhus C\\
              Denmark
}

\date{Received: date / Accepted: date}

\maketitle

\begin{abstract}
We consider renormalization of effective field theory interactions by discretizing the continuum on a tight-binding lattice. After studying the one-dimensional problem, we address s-wave collisions in three dimensions and relate the bare lattice coupling constants to the continuum coupling constants. Our method constitutes a very simple avenue for the systematic renormalization in effective field theory, and is especially useful as the number of interaction parameters increases.
\end{abstract}

\section{Introduction}
The physics of quantum systems with strong interparticle interactions 
is under intense study on several 
fronts as we are currently experiencing a convergence of different areas and methods 
that hold promise to yield important insights into this crucial problem. Experiments
at RHIC and LHC are expected to provide new details on the strong interaction dynamics of 
nuclear and particle physics \cite{braun2012}, while advances in atomic, molecular and 
optical physics have opened up the regime of strong interactions for many-body physics
with atoms and molecules \cite{bloch2008,chin2010}. The latter may be used to mimick 
the conditions found in strongly-interacting condensed-matter environments \cite{lewenstein2007,esslinger2010}, or even 
neutron stars \cite{Baker,Pethick,Carlson}.
There is great hope that these
developments will complement and cross-fertilize each other in coming years \cite{zinner2013}.

The theoretical study of systems with strong interactions can be extremely complicated as
any na{\"i}ve perturbative approach is doomed to fail. However, the tremendous increase in 
computational power and the development of new methods means that one can now perform
simulations on discretized lattices of strongly-interacting few- and many-body dynamics 
that are not only extremely precise
but also yield controlled estimations of uncertainties \cite{lee2009,endres2013}. The results of 
such simulations can subsequently be extrapolated to the limit of zero lattice spacing
and predictions for continuum dynamics may be obtained. To utilize these powerful simulation
tools it is crucial to have a fully consistent description of two-body interaction 
terms on a lattice, and furthermore be able to connect this knowledge to parameters
that are often defined and measured in free space. Examples are the scattering length
or effective range parameters that describe low-energy scattering and have a long 
history in nuclear and atomic physics.

Within the realm of effective field theory (EFT), the simulation techniques discussed above
are extremely useful for studying low-energy dynamics in the presence of strong
interactions. EFT originates from Weinberg's expansion of the nucleon-nucleon 
potential at low momenta (compared to the nucleon mass) with chiral perturbation 
theory \cite{weinberg1990}. There are issues with regularization and renormalization 
of this expansion that have to be treated with great care, particularly when 
higher-order terms are included \cite{phillips1998,kaplan1998a,kaplan1998b}. Furthermore,
these issues become more intricate on a lattice as discussed in the seminal work 
of L{\"u}scher \cite{luscher1986}.

In this paper we consider regularization and renormalization of two-body interactions 
in EFT and provide some solvable, physically important, and technically simpler models than those 
that are typically discussed. We start in one-dimensional settings where the 
procedures are particularly clear, and so far unexplored. This is done for both $s$- and $p$-wave effective 
interactions on a lattice for which the continuum limit may be performed in 
the end. One-dimensional systems are realizable with cold atoms \cite{bloch2008} and
strongly interacting one-dimensional gases have been produced in 
recent years to study both many-body \cite{paredes2004,kinoshita2004,kinoshita2005,haller2009}
and few-body physics \cite{serwane2011,zurn2012}. After discussing the one-dimensional 
renormalization on the lattice, we proceed to the three-dimensional case and demonstrate
how the lessons learned in one dimension may be used, at zero extra cost, to renormalize the interactions in 
a much more straightforward manner than the direct approach via the Lippmann-Schwinger equation. We also provide the direct link between the discretized and continuum EFT parameters, especially important for the three-dimensional case.

\section{Effective field theory}
In effective field theory (EFT) of scattering, the momentum representation of the two-body interaction, $V(\mathbf{k},\mathbf{k}')$, is expanded in series as
\begin{equation}
V(\mathbf{k},\mathbf{k}') = g_0+\frac{g_2}{2}(\mathbf{k}^2+\mathbf{k}'^2)+g_{11}\mathbf{k}\cdot \mathbf{k}' + \ldots .\label{EFTexpansion}
\end{equation}
In three spatial dimensions (3D), the above interaction is renormalizable, at least to order $k^2$, and yields the correct s-wave scattering amplitude or, equivalently, the phase-shifts, at low energy. That is, the effective range expansion
\begin{equation}
k\cot(\delta_0)=-\frac{1}{a}+\frac{r}{2}k^2+\ldots,
\end{equation}
where $a$ and $r$ are, respectively, the s-wave scattering length and effective range, while $\delta_0$ is the s-wave phase-shift. 

The renormalization of the interaction in Eq. (\ref{EFTexpansion}) has been carried out by solving the Lippmann-Schwinger equation for the T-matrix \cite{Cohen,frederico2000,Braaten}. This procedure, while conceptually simple, disembarks in frustratingly tedious algebra which masks the physics of the problem considerably. We here address the problem by discretizing the Schr\"odinger equation on a lattice. This method largely simplifies the problem and allows a clear interpretation of the results. Before considering the three-dimensional problem, we study one-dimensional scattering for which the renormalization prescription has not been elucidated yet. We also solve the problem of 1D ``p-wave'' interactions, where universal single-parameter renormalization does work \cite{Cheon,Muth}, in contrast with the 3D case where it does not \cite{Gurarie,Levinsen}.

\section{One-dimensional collisions}
We define a 1D tight-binding lattice with two particles that can occupy discrete positions $x=jd$, where $j$ is an integer and $d$ is the lattice spacing. The lattice kinetic energy dispersion is chosen to be 
\begin{equation}
E(k)=-2J\cos(kd)+2J, \label{energydispersion}
\end{equation}
which is quadratic in $k$ close to the continuum limit ($d\to 0$). In momentum space, an interaction with a low-energy Taylor series given by Eq. (\ref{EFTexpansion}) while respecting the periodicity of the Brillouin zone can be chosen as 
\begin{equation}
V(k,k')/d=U+2V\cos[(k-k')d].\label{latticeinteraction}
\end{equation}

\subsection{S-wave interaction}
We first handle the s-wave part of the EFT interaction, that is, the renormalization of the coupling constants $g_0$ and $g_2$.
The lattice Hamiltonian corresponding to the energy dispersion (\ref{energydispersion}) and the interaction (\ref{latticeinteraction}) is given by
\begin{equation}
-J\sum_{j}(b^{\dagger}_{j+1}b_j+b^{\dagger}_jb_{j+1})+\sum_j \left[\frac{U}{2}n_j(n_j-1)+Vn_jn_{j+1}+2Jn_j\right].\label{HamiltonianUV}
\end{equation}
Above, $b_j$ ($b^{\dagger}_j$) is the bosonic annihilation (creation) operator at the lattice site $j$, and $n_j=b^{\dagger}_jb_j$ is the number operator. We have chosen to work with bosons for simplicity, but the following applies to spin-$1/2$ fermions, too. It must be noted that, with fermions, the interaction in Eq. (\ref{latticeinteraction}) must be made spin-dependent, for otherwise the s-wave renormalized interaction would imply a hard-core ``p-wave'' interaction for triplet fermions, as we will see. 

The two-body phase-shift for the Hamiltonian (\ref{HamiltonianUV}) is given by \cite{ValienteJPB2009}
\begin{equation}
-\sin(kd)\cot \delta_0 = \frac{2u+\left[4\cos(kd)+u\right]v\cos(kd)}{uv-4\left[2-v\cos(kd)\right]},\label{phaseshift1}
\end{equation}
where we have defined the dimensionless constants $u=U/J$ and $v=V/J$. To take the continuum limit, we expand the trigonometric functions of $kd$ to second order in $kd$. The phase-shift therefore becomes
\begin{equation}
-k\cot \delta_0 = \frac{1}{d}\frac{2u+4v+uv-vk^2d^2(4+u)/2}{uv+4v-8}.
\end{equation}
Above, we have assumed, in order not to have any $k^2$-dependence in the denominator, that $vd^2\to 0$ as $d\to 0$. We will see, after renormalization, that the assumption is indeed correct. Using the 1D effective range expansion, $-k\cot\delta_0 = \alpha+\beta k^2$, where $\alpha=1/a$ and $\beta=-r$, we find
\begin{align}
u&=-4\frac{\alpha d^2+2\beta}{d+2\beta}\\
v&=4\frac{\beta}{d-d^2\alpha}.\label{nearestneighbor}
\end{align}
Close to the continuum limit we have $v\sim 1/d$, and therefore our assumption that $vd^2\to 0$ is correct.

We now proceed to relate the lattice parameters $u$ and $v$ to the continuum EFT bare coupling constants $g_0$ and $g_2$ of Eq. (\ref{EFTexpansion}). Close to the continuum limit, the lattice potential (\ref{latticeinteraction}) is given by
\begin{equation}
V(k,k')/d=U+2V-V[(k-k')d]^2+O[(k-k')^4d^4].\label{Vlatticecont}
\end{equation}
We then equate Eq. (\ref{EFTexpansion}) to Eq. (\ref{Vlatticecont}) in the s-wave sector, obtaining
\begin{align}
U+2V&=g_0/d\\
-Vd^2&=\frac{g_2}{2d}.
\end{align}
The tunnel coupling constant $J$ close to the continuum limit has to be adjusted in such a way that the energy dispersion, Eq. (\ref{energydispersion}) is given by $\hbar^2k^2/2m$. Therefore, $J$ has the form 
\begin{equation}
J=\frac{\hbar^2}{2md^2}
\end{equation}
as $d\to 0$. The explicit expressions for the continuum coupling constants are given by
\begin{align}
g_0&=\frac{\hbar^2}{2md}\left[\frac{8\beta}{d-d^2\alpha}-4\frac{d^2\alpha+2\beta }{d+2\beta}\right]\\
g_2&=-4\frac{\hbar^2\beta}{m(1-d\alpha)}.
\end{align}
The behavior of the coupling constants to leading order is easily recognized as $g_0\sim d^{-2}+O(d^{-1})$ and $g_2\sim 1$. 

\subsection{P-wave interactions}
We now renormalize the p-wave coupling constant $g_{11}$. This problem, which is simpler than the for s-wave interactions, has already been solved (see \cite{Muth}) by comparing the action of the finite-difference kinetic energy operator on a fermionic wave function with Cheon-Shigehara boundary conditions \cite{Cheon} in the continuum. 

We recover the renormalization result for $g_{11}$, and give a simple physical interpretation here by using the phase-shift in Eq. (\ref{phaseshift1}), which becomes the phase-shift for two polarized fermions in the limit $U\to \infty$ \cite{ValienteJPB2009}, in virtue of the Bose-Fermi mapping theorem \cite{Girardeau}. Close to the continuum limit, the phase-shift becomes
\begin{equation}
-k\cot \delta_1 = \frac{1}{d}\frac{2+v(1-k^2d^2/2)}{v}.
\end{equation}
Matching the relation above to the effective range expansion
\begin{equation}
-k\cot\delta_1 =\alpha+\beta k^2,
\end{equation}
we obtain $\beta=-d/2$, and 
\begin{equation}
v=-\frac{2}{1-d\alpha}.
\end{equation}
Interestingly, setting $\beta=-d/2$ implies $u\to \infty$, too. Therefore, we can allow interactions to be spin-independent for hard-core spin-$1/2$ fermions with nearest-neighbor interaction. This, on the other hand, would imply a magical matching of s-wave and p-wave scattering amplitudes and therefore, even in the hard-core case, spin-dependent interactions would be strongly preferred based on physical grounds. In the case of finite $u$, however, the nearest-neighbor interaction, Eq. (\ref{nearestneighbor}), diverges as $v\sim d^{-1}$ and spin-independent interactions would lead to a nearest-neighbor hard-core condition for pairs  of triplet fermions. 

Finally, the bare coupling constant $g_{11}$ is renormalized as 
\begin{equation}
g_{11}=\frac{\hbar^2vd}{m}=-\frac{2\hbar^2}{m}\frac{d}{1-d\alpha},
\end{equation}
which vanishes as $d\to 0$, but its dependence on the lattice spacing through $d\alpha$ must be kept in all calculations before taking the continuum limit. This is the exactly same situation we observe in the single-parameter EFT in three-dimensions, and we will see in the next section that these two problems are formally equivalent. Clearly, a linearly vanishing coupling constant $g_{11}$ tells us that the bare continuum limit has a linear UV divergence.

We now consider the next-to-leading order correction for p-wave scattering in 1D. This problem, which is relevant {\it per se}, can actually be used to renormalize the s-wave bare EFT parameters $g_0$ and $g_2$ in three dimensions, as we will show in the next section. 

To solve the problem, we first invoke the Bose-Fermi mapping theorem so that we do not need to work with fermions. The Hamiltonian $H'$ of the system is thus given by Eq. (\ref{HamiltonianUV}) with $U\to \infty$ and an extra term, as
\begin{equation}
H'=H+W\sum_{j}n_jn_{j+2}.\label{fermiHam}
\end{equation}
The phase-shift is easily obtained by using the methods of \cite{ValientePRA2010} for arbitrary finite-range interactions on a 1D lattice. Defining $v=V/J$ and $w=W/J$, the phase-shift reads
\begin{equation}
-\cot\delta_1=\frac{4+2(v+w)\cos(kd)+vw\cos(2kd)+2w\cos(3kd)}{2(v+2w+vw\cos(kd)+2w\cos(2kd))\sin(kd)}.\label{phaseshiftfermi}
\end{equation}
Assuming now that $(vw+2w)d^2\to 0$ as we approach the continuum limit, and fitting $v$ and $w$ to match the effective range expansion, we obtain
\begin{align}
\alpha&=\frac{1}{d}\frac{4+2v+4w+vw}{2v+8w+2vw}\\
\beta&=-d(v+2vw+10w),
\end{align}
The lattice interaction parameters can have two different values each, denoted as $v_{\pm}$ and $w_{\pm}$, given by
\begin{align}
v_{\pm}&=\frac{1}{2d(2d\alpha-3)}\left(24d-12d^2\alpha-\beta+2d\alpha\beta\pm \Gamma\right)\label{v}\\
w_{\pm}&=\frac{1}{4d(2d\alpha-1)}\left(8d-12d^2\alpha+\beta-2d\alpha\beta\pm \Gamma\right)\label{w}
\end{align}
where we have defined
\begin{equation}
\Gamma=\sqrt{16d(1-2d\alpha)[d(2+\alpha\beta)-\beta]+[\beta-2d(6d\alpha+\alpha\beta-4)]^2}.
\end{equation}
We can now see that $(v_{\pm}w_{\pm}+2w_{\pm})=(\beta\pm |\beta|)d+O(d^2)$, and therefore our assumption was correct.
The above expressions for the lattice coupling constants is now more complicated than in the previous cases. We will see how this is completely equivalent to three-dimensional s-wave EFT when the parameters are properly re-defined in the following section.

\section{Three-dimensional collisions}
In three spatial dimensions, there is a rather intrincate relation \cite{Cohen,Braaten} between the bare EFT parameters $g_0$ and $g_2$ and the scattering length $a$ and effective range $r$ to renormalize the two-body problem in the s-wave sector. Typically, this relation is obtained by solving the corresponding Lippmann-Schwinger equation non-perturbatively and, as we mentioned earlier, involves rather cumbersome and tedious algebraic manipulations. We here use the results of the previous section to obtain these relations in a straightforward way.

We begin with the reduced radial Hamiltonian in the s-wave channel 
\begin{equation}
H=-\frac{\hbar^2}{2\mu}\frac{d^2}{dr^2}+V(r).\label{radialHam}
\end{equation}
The eigenfunctions of $H$ are the reduced radial wave functions $u(r)=rR(r)$, where $R(r)$ are the radial wave functions of the original radial Hamiltonian. The discretization of $H$ is trivial, and we only need to implement the boundary condition $u(r<0)=0$ for the reduced wave functions. A little care must be taken in using Dirac delta functions now. To illustrate this, we look at the bare delta interaction $V_{D}$, which has the form
\begin{equation}
V_{D}(r)=g_0\frac{\delta(r)}{4\pi r^2}.
\end{equation}
Direct discretization of the above interaction, if we allow the point $j=0$ to be included on the lattice, is ambiguous. However, $r=jd\to 0$ for any finite $j$. Therefore, we can simply define the lattice coordinate space for $j\ge 1$ without loss of generality, and implement the boundary condition $u(0)=0$ before the continuum limit is taken. The discretized version of the Hamiltonian (\ref{radialHam}) with nearest and next-nearest neighbor interactions on the lattice is therefore exactly given by the 1D hard-core (or fermionic) Hamiltonian of Eq. (\ref{fermiHam}). The phase-shifts are given by Eq. (\ref{phaseshiftfermi}). The only difference now lies in adjusting the EFT bare parameters $g_0$ and $g_2$ for the three-dimensional case.

The Fourier transform of the s-wave part of the interaction in continuous space is given by
\begin{equation}
V_S(q)=4\pi\int_{0}^{\infty}dr r^2 j_0(qr) V(r),
\end{equation}
where $q=|\mathbf{k}-\mathbf{k}'|$, and $j_0(x)=\sin(x)/x$ is the zero-th order spherical Bessel function. We take the low-energy expansion of $V_S$ on the left hand side, $V_S(q)=g_0+g_2q^2/2+O(q^4)$. Then, we discretize the Fourier transform for our lattice problem (note that this is not the same as using the discrete Fourier transform) using Riemann's sums, as
\begin{equation}
V_{\mathrm{lattice}}(q) = 4\pi d^3\left[Vj_0(qd)+4Wj_0(2qd)\right].
\end{equation}
Expanding the above expression to fourth order in $(kd)$, and equating the resulting expansion to the continuum limit, we find
\begin{align}
g_0&=4\pi\left(V+4W\right)d^3,\label{g03D}\\
\frac{g_2}{2}&=-4\pi\left(\frac{V}{6}+\frac{W}{3}\right)d^5,\label{g23D}\\
&d^7\left(\frac{V}{120}+\frac{8W}{15}\right)\to 0.
\end{align}
The effective range expansion in three-dimensions looks exactly like its one-dimensional counterpart, and therefore the lattice coupling constants are here given by Eqs. (\ref{v}) and (\ref{w}). The EFT coupling constants $g_0$ and $g_2$ follow immediately from Eqs. (\ref{v},\ref{w},\ref{g03D},\ref{g23D}). 

The only point we have to address now, in order to be able to use the EFT interaction in continuous space, is to relate the lattice spacing $d$ to the cut-off $\Lambda$. Note that this is unnecessary in 1D, where we can do all relevant calculations on the lattice at the $N$-body level. We can set $W=0$ and obtain the bare coupling constant $g_0$ as a function of the lattice spacing and the scattering length $a=1/\alpha$. We obtain 
\begin{equation}
g_0=\frac{2\pi \hbar^2d}{m} \frac{2}{d/a-1}.\label{g01}
\end{equation}
From standard cut-off renormalization, we have
\begin{equation}
g_0 = \frac{4\pi \hbar^2 a}{m} \frac{1}{1-2a\Lambda/\pi}.\label{g02}
\end{equation}
Comparing now Eqs. (\ref{g01}) and (\ref{g02}), we see that $d=\pi/2\Lambda$, which is the desired relation.

\section{Conclusions and outlook}
We have presented a simple way of renormalizing two-body effective field theory (EFT) interactions by using a discrete lattice and taking the correct continuum limit. In one dimension, the lattice discretization can be used for the many-body problem, where numerically accurate methods, such as the density matrix renormalization group (for a review, see \cite{DMRG}) are available. Therefore, our renormalization prescriptions can be used to obtain effective-range corrections in an essentially exact fashion. In the three-dimensional case, we discretized the s-wave radial Schr\"odinger equation, and used the equivalent one-dimensional results to obtain the EFT coupling constants in a straightforward manner.

Our results can be extended to include a finite, arbitrary number of coupling constants \cite{ValientePRA2010}, and may be useful to clarify, whithout the tedious intricacies of the multichannel Lippmann-Schwinger equation in continuous space, whether quartic s-wave interactions in three dimensions are renormalizable \cite{Braaten,Yang}. Extension of our method to higher angular momentum channels may be possible by using a mixed lattice-continuous representation of the partial-wave radial Schr\"odinger equation. Of special relevance is the problem of two-parameter renormalization of p-wave interactions, which is important for highly polarized Fermi gases \cite{Chevy,Levinsen} and p-wave superfluids \cite{Gurarie}. 



\begin{thebibliography}{3}
%
%

\bibitem{braun2012} J. Braun, J. Phys. G {\bf 39}, 033001 (2012).
\bibitem{bloch2008} I. Bloch, J. Dalibard, and W. Zwerger, Rev. Mod. Phys. {\bf 80}, 885 (2008).
\bibitem{chin2010} C. Chin, R. Grimm, P. Julienne, and E. Tiesinga, Rev. Mod. Phys. {\bf 82}, 1225 (2010).
\bibitem{lewenstein2007} M. Lewenstein {\it et al.}, Adv. Phys. {\bf 56}, 243 (2007).
\bibitem{esslinger2010} T. Esslinger, Ann. Rev. Cond. Mat. Phys. {\bf 1}, 129 (2010).
\bibitem{Baker}
G.~A. Baker, Jr.,
Phys. Rev. C {\bf 60}, 054311 (1999).
\bibitem{Pethick}
A. Schwenk and C.~J. Pethick,
Phys. Rev. Lett. {\bf 95}, 160401 (2005).
\bibitem{Carlson}
J. Carlson, J. Morales, Jr., V.~R. Pandharipande and D.~G. Ravenhall,
Phys. Rev. C {\bf 68}, 025802 (2003).
\bibitem{zinner2013} N.~T. Zinner and A.~S. Jensen, J. Phys. G: Nucl. Part. Phys. {\bf 40}, 053101 (2013).
\bibitem{lee2009} D. Lee, Prog. Part. Nucl. Phys. {\bf 63}, 117 (2009).
\bibitem{endres2013} M.~G. Endres, D.~B. Kaplan, J.-W. Lee, and A.~M. Nicholson, Phys. Rev. A {\bf 87}, 023615 (2013).
\bibitem{weinberg1990} S. Weinberg, Phys. Lett. B {\bf 251}, 288 (1990); Nucl. Phys. B {\bf 363}, 3 (1991).
\bibitem{phillips1998} D.~R. Phillips, S.~R. Beane, and T.~D. Cohen, Nucl. Phys. A {\bf 631}, 447 (1998).
\bibitem{kaplan1998a} D.~B. Kaplan, M.~J. Savage, and M.~B. Wise, Phys. Lett. B {\bf 424}, 390 (1998).
\bibitem{kaplan1998b} D.~B. Kaplan, M.~J. Savage, and M.~B. Wise, Nucl. Phys. B {\bf 534}, 329 (1998).
\bibitem{luscher1986} M. L{\"u}scher, Commun. Math. Phys. {\bf 104}, 177 (1986); 
Commun. Math. Phys. {\bf 105}, 153 (1986); Nucl. Phys. B {\bf 354}, 531 (1991).
\bibitem{paredes2004} B. Paredes {\it et al.}, Nature (London) {\bf 429}, 277 (2004).
\bibitem{kinoshita2004} T. Kinoshita, T. Wenger, and D. Weiss, Science {\bf 305}, 1125 (2004).
\bibitem{kinoshita2005} T. Kinoshita, T. Wenger, and D. Weiss, Phys. Rev. Lett. {\bf 95}, 190406 (2005).
\bibitem{haller2009} E. Haller {\it et al.}, Science {\bf 325}, 1224 (2009).
\bibitem{serwane2011} F. Serwane {\it et al.}, Science {\bf 332}, 336 (2011).
\bibitem{zurn2012} G. Z{\"u}rn {\it et al.}, Phys. Rev. Lett. {\bf 108}, 075303 (2012).



\bibitem{Cohen}
D.~R. Phillips, S.~R. Beane and T.~D. Cohen,
Ann. Phys. (NY) {\bf 263}, 255 (1998).

\bibitem{frederico2000}
T. Frederico, A. Delfino and L. Tomio,
Phys. Lett. B {\bf 481}, 143 (2000).

\bibitem{Braaten}
E. Braaten, M. Kusunoki, D. Zhang,
Ann. Phys. (NY) {\bf 323}, 1770 (2008).

\bibitem{Cheon}
T. Cheon and T. Shigehara,
Phys. Rev. Lett. {\bf 82}, 2536 (1999).

\bibitem{Muth}
D. Muth, M. Fleischhauer and B. Schmidt,
Phys. Rev. A {\bf 82}, 013602 (2010).

\bibitem{Gurarie}
V. Gurarie and L. Radzihovsky,
Ann. Phys. (NY) {\bf 322}, 2 (2007).

\bibitem{Levinsen}
J. Levinsen, P. Massignan, F. Chevy and C. Lobo,
Phys. Rev. Lett. {\bf 109}, 075302 (2012).

\bibitem{ValienteJPB2009}
M. Valiente and D. Petrosyan, J. Phys. B: At. Mol. Opt. Phys. {\bf 42}, 121001 (2009).

\bibitem{Girardeau}
M.~D. Girardeau,
J. Math. Phys. (1960).

\bibitem{ValientePRA2010}
M. Valiente,
Phys. Rev. A {\bf 81}, 042102 (2010). 

\bibitem{DMRG}
K.~A. Hallberg,
Adv. Phys. {\bf 55}, 477 (2010).



\bibitem{Yang}
J.-F. Yang and J.-H. Huang,
Phys. Rev. C {\bf 71}, 034001 (2005). 

\bibitem{Chevy}
F. Chevy and C. Mora,
Rep. Prog. Phys. {\bf 73}, 112401 (2010).
 

\end{thebibliography}


\end{document}